\documentclass{iopconfser}
\usepackage{soul,todonotes}

\usepackage{amsmath,amssymb,mathrsfs}

\usepackage{tikz}

\newcommand{\CC}{\mathbb{C}}

\newcommand{\Af}{{\mathscr A}}
\newcommand{\Bf}{{\mathscr B}}
\newcommand{\Cf}{{\mathscr C}}

\newcommand{\Uf}{{\mathscr U}}
\newcommand{\Zf}{{\mathscr Z}}

\newcommand{\ee}{\mathrm{e}}
\newcommand{\ii}{\mathrm{i}}
\newcommand{\dvol}{\mathrm{dvol}}
\newcommand{\rfhgho}{RFHGHO}
\newcommand{\one}{\boldsymbol{1}}

\DeclareMathOperator{\supp}{supp}

\begin{document}

\title{Measurement and preparation protocols for quantum field theory on curved spacetimes}

\author{Christopher J.~Fewster$^{1,2}$}

\affil{$^1$Department of Mathematics, Ian Wand Building, Deramore Lane, University of York, York YO10 5GH, UK}\bigskip
  
\affil{$^2$York Centre for Quantum Technologies, University of York, York YO10 5DD, UK}

\email{chris.fewster@york.ac.uk}

\begin{abstract}
	In this conference proceedings contribution, I describe work in progress concerning
	two problems in the measurement theory of quantum fields. 
	First, it is proved that all local observables can be obtained from local measurement schemes. Second, I describe a protocol for preparing a Hadamard local product state of given Hadamard states relative to the local algebras of specified spacelike separated regions. 
\end{abstract}

\section{Introduction}

This contribution concerns the measurement of local observables in quantum field theory (QFT) on curved spacetimes and the ability of spacelike separated experimenters to prepare independent physically reasonable states for measurements on a given QFT, working within a framework 
introduced in~\cite{FewVer_QFLM:2018} (see~\cite{FewsterVerch_encyc:2023} for a summary).
A full version of the results presented here will appear in~\cite{Fewster:202?}.

Briefly, in the framework of~\cite{FewVer_QFLM:2018}, one considers two quantum field theories on spacetime $M$, called the \emph{system} and \emph{probe}, whose algebras of local observables are labelled $\Af$ and $\Bf$ respectively. Their tensor product defines the \emph{uncoupled combination} $\Uf$; in addition, we consider a modified \emph{coupled combination} $\Cf$ in which $\Af$ and $\Bf$ are coupled in some temporally compact spacetime subset $K$. The effect of this coupling can be captured in a \emph{scattering map} $\Theta$, which is an automorphism on $\Uf$. At early times, the system and probe are prepared independently in states $\omega$ and $\sigma$. It is shown in~\cite{FewVer_QFLM:2018} that, a late time measurement of probe observable $B$ provides an indirect measurement of the \emph{induced system observable}
\begin{equation}
	\varepsilon_\sigma(B):=\eta_\sigma(\Theta (\one\otimes B)),
\end{equation}
where $\eta_\sigma:\Uf\to\Af$ is defined so that $\eta_\sigma(A\otimes B)=\sigma(B)A$. 
Collectively, the probe $\Bf$, coupled theory $\Cf$, probe preparation state $\sigma$ and probe observable $B$ constitute a \emph{measurement scheme} for $	\varepsilon_\sigma(B)$.
Furthermore, one can determine an updated state that can be used for further predictions following either a selective or nonselective measurement of a binary observable $B$. The nonselective update, for example, is given by
\begin{equation}
	\omega_{\text{ns}}(A)= (\omega\otimes\sigma)(\Theta(A\otimes \one)).
\end{equation}

Numerous good properties of this framework are discussed in~\cite{FewVer_QFLM:2018,BostelmannFewsterRuep:2020,FewsterJubbRuep:2022}. In particular, it is free of the `impossible measurement' problem noted by Sorkin~\cite{sorkin1993impossible} as shown in~\cite{BostelmannFewsterRuep:2020}. It has been shown recently that the nonselective update of a Hadamard state is itself Hadamard
(assuming the probe state is also Hadamard) within a class of theories described by real formally hermitian Green-hyperbolic operators (\rfhgho's) obeying a decomposability condition~\cite{Fewster:2025a}. \rfhgho's are a very natural class of operators,  describing the Klein--Gordon field and similar normally hyperbolic models for fields valued in vector bundles, but also non-hyperbolic models such as the Proca field. By a doubling trick, complex fields -- such as the charged Klein--Gordon theory in an external electromagnetic field -- may be treated, and fermionic fields may be incorporated by adding additional structure. Assuming the decomposability condition mentioned above, a generalised Hadamard condition may be formulated that extends many of the standard properties of Hadamard states and is stable under operations such as tensor product, partial trace and the action of scattering morphisms, all of which is described in~\cite{Fewster:2025a}.
 
The measurement framework of~\cite{FewVer_QFLM:2018} has also provided a basis for an investigation of measurement on spacetimes with symmetries~\cite{FewsterJanssenLoveridgeRejznerWaldron:2024}, generalising~\cite{CLPW:2023} and finding the same phenomenon of a change in type of local von Neumann algebras.

Here, we consider two further developments of this framework, 
concerning the design of measurement schemes and the ability of experimenters to prepare desired states, and answering the following questions:
\begin{enumerate}
	\item Can one find a measurement scheme for every local system observable? 
	\item Two experimenters, the ubiquitous Alice and Bob, who wish to perform 
	independent local measurements on the system QFT in spacelike separated spacetime regions\footnote{A region is an open, causally convex spacetime subset; in particular, it is globally hyperbolic in its own right.} $O_A$ and $O_B$. Alice wants to measure in state $\omega_A$, while Bob wants to measure in state $\omega_B$. Can this dilemma be resolved? 
\end{enumerate}
The first question has been partly addressed before. On one hand, it has been shown that any local observable can be realised as the limit of a sequence of induced observables -- a so-called \emph{asymptotic measurement scheme}~\cite{FewsterJubbRuep:2022}. On the other hand, restricting to a class of Gaussian modulated POVMs, there are simple formulae relating the parameters of the probe POVM to the POVM induced in the system; inverting these formulae, one obtains an exact solution to problem 1 for this class~\cite{MandryschNavascues:2024}. However, it remains to show that \emph{all} local observables can be obtained exactly from a measurement scheme with a local coupling. 

The second question would be (nearly) resolved if there is a state $\omega$ of the system theory so that 
\begin{equation}
	\omega(XY) = \omega_A(X)\omega_B(Y) ,\qquad \forall~ X\in\Af(O_A), \quad Y\in \Af(O_B),
\end{equation}
where $\Af(O_{A/B})$ are the subalgebras of observables localisable in Alice/Bob's laboratory. One sees that the expectation values of any observable in Alice's lab in state $\omega$ agree with those in Alice's desired state $\omega_A$, and likewise for Bob. Importantly, $\omega$ has no correlations between observables in $\Af(O_A)$ and $\Af(O_B)$, so the experiments produce independent results. A state with the above property is called a \emph{local product state}. To fully resolve the second question we require that there exists a local product state of $\omega_A$ and $\omega_B$ that is physically reasonable, given that $\omega_A$ and $\omega_B$ are physically reasonable. While there are general existence results for local product states 
(see~\cite{Summers:2009} for a review on this and the wider question of independence of subsystems) these do not give much control over the state itself, and with a rather weak definition of physical reasonableness. More recently, Sanders~\cite{Sanders_separable:2023} has shown the existence of Hadamard states with a local product form for the Klein--Gordon field in curved spacetimes. However, this does not construct local product states from given $\omega_A$, $\omega_B$, but rather proceeds by constructing a local product state in Minkowski spacetime and then using deformation methods to obtain states with this property in general globally hyperbolic spacetimes.

We will provide positive resolutions to both questions. For simplicity, we restrict the presentation here to the real Klein--Gordon field, but the ideas can be generalised to a wide class of theories described by \rfhgho's.

\section{Gauge symmetry from scattering}

The heart of the construction is based on the following observations. Let $\Af$ 
represent the theory of a Klein--Gordon field of mass $m\ge 0$ on a globally hyperbolic spacetime $M$ with mostly negative signature. Then $\Af\otimes\Af$ represents two fields of the same mass, and can be regarded as the uncoupled combination of a system and probe which are copies of the Klein--Gordon field. However $\Af\otimes\Af$ can be identified with the QFT $\Zf$ representing a single complex Klein--Gordon field of mass $m$. $\Zf$ is a unital $*$-algebra generated by smeared fields $Z(f)$ and $\overline{Z}(h)$, where $f$ and $h$ are compactly supported smooth functions on $M$, subject to the following conditions
\begin{itemize}
	\item the maps $f\mapsto Z(f)$ and $h\mapsto \overline{Z}(h)$ are complex-linear
	\item $Z(f)^*=\overline{Z}(\overline{f})$
	\item $Z((\Box+m^2)f)=0= \overline{Z}((\Box+m^2)h)$
	\item $[Z(f),Z(h)]=[\overline{Z}(f),\overline{Z}(h)]=0$, $[Z(f),\overline{Z}(h)]=\ii
	E(f,h)\one$
\end{itemize}  
for all test functions $f,h$. Here, $E(f,h)=\int f(x) (E^-(x,y)-E^+(x,y))h(y)\dvol(x)\dvol(y)$, where $E^{{+/-}}$ are the retarded/advanced Green functions for the Klein--Gordon operator.
There is a $\mathrm{U}(1)$ global gauge symmetry: for each $z\in\CC$ with $|z|=1$ there is an automorphism $\alpha(z)$ of $\Zf$ defined uniquely by its action 
\begin{equation}
	\alpha(z) Z(f) = Z(zf), \qquad \alpha(z) \overline{Z}(f)= \overline{Z}(\bar{z}f)
\end{equation}
on smeared fields. The key to our approach is that every $\alpha(z)$ can be obtained as a scattering map obtained from a coupled version of $\Zf$. Choose Cauchy surfaces $\Sigma^\pm$ in $M$ so that $\Sigma^+$ lies strictly to the future of $\Sigma^-$. Then we may choose a smooth function $\chi:M\to\mathrm{U}(1)$ so that $\chi\equiv 1$ to the future of $\Sigma^+$ and $\chi\equiv z$ to the past of $\Sigma^-$. Writing $P=\Box+m^2$, let
\begin{equation}
	Q = \chi P \chi^{-1}  =  (\nabla^\mu + \ii A^\mu) (\nabla_\mu+ \ii A_\mu), \qquad A=-\nabla\chi.
\end{equation}
Clearly $Q$ is the theory of a complex Klein--Gordon field in an external pure gauge electromagnetic potential $A$, which vanishes outside the spacetime slab enclosed by $\Sigma^\pm$. Taking $\Cf$ to be the QFT of this field, one may show that the scattering map $\Theta_z$ of $\Cf$ relative to $\Zf=\Af\otimes\Af$ is precisely the automorphism $\alpha(z)$. Note that $\Cf$ depends on the function $\chi$, but $\Theta_z$ depends only on $z$.  

\section{Measurement schemes and state preparation -- noncompact coupling zone}
\label{sec:noncompact}

The $\mathrm{U}(1)$ symmetry of the complex theory corresponds to a global $\mathrm{SO}(2)$ symmetry of the two real Klein--Gordon fields, relabelling $z=\ee^{\ii s}$. In particular, for $s=\pm\pi/2$ it turns out that the scattering map $\Theta_s$ acts in a simple way on the observables of $\Af\otimes\Af$, namely
\begin{equation}
\Theta_{\pi/2} A\otimes\one = \one\otimes A, \qquad \Theta_{-\pi/2} \one\otimes A = A\otimes 1
\end{equation}
The expression for the induced observable map $\varepsilon_\sigma$ when $s=-\pi/2$ is then
\begin{equation}\label{eq:vareps_id}
\varepsilon_{\sigma}(B) = \eta_\sigma(\Theta_{-\pi/2}(\one\otimes B)) = \eta_\sigma(B\otimes\one) =B,
\end{equation}
while the formula for the nonselectively updated state for a coupling with $s=\pi/2$ gives $\omega_{\text{ns}}=\sigma$, because
\begin{equation}\label{eq:nonsel_id}
	\omega_{\text{ns}}(A) = (\omega\otimes\sigma)(\Theta_{\pi/2}( A\otimes \one)) = (\omega\otimes\sigma)(\one\otimes A) =\sigma(A).
\end{equation}
These expressions show that every observable of $\Af$ has a measurement scheme, and every state can be prepared (at least, assuming it can be prepared on the probe). For measurement schemes, this argument appears as a `proof of principle' in~\cite{FewsterJubbRuep:2022}, but it is not satisfactory as an answer to our questions because it relies on controlling the whole slab between $\Sigma^\pm$, which is experimentally infeasible.

\section{Measurement schemes and state preparation -- compact coupling zone}

The idea needed to obtain more physically reasonable versions of the measurement scheme and preparation protocols is a localisation property of the measurement framework that will appear in work with Benito Ju\'arez-Aubry~\cite{FewsterJuarezAubry:202?}. In outline, this shows the following: consider spacetime regions $L^\pm$ and $L$ so that
$L^+\cup L^-\subset L\subset D^+(L^-)$, where $D^+$ denotes the future Cauchy development, so that $L^+$ ($L^-$) lies to the future (past) of a spacetime slab $S$.
Consider two coupled theories $\Cf$ and $\Cf'$ with coupling zones in $S$
and suppose they coincide on the region $L$; consider probe preparation states $\sigma$ and $\sigma'$ that agree on $\Bf(L^-)$; consider system states $\omega$ and $\omega'$ that agree on $\Af(L^-)$. Then the induced observable maps determined by $(\Cf,\sigma)$ and $(\Cf',\sigma')$ agree on $\Bf(L^+)$ and the nonselective updates of $\omega$ and $\omega'$ agree on $\Af(L^+)$. (Some details are suppressed and in fact more is shown in~\cite{FewsterJuarezAubry:202?}.)   

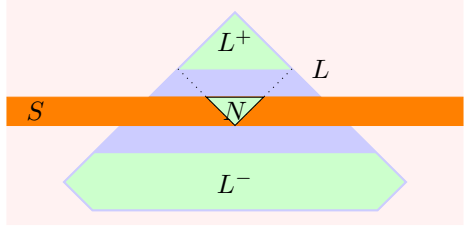
\begin{figure}
		\begin{center}
		\begin{tikzpicture}[scale=0.75]
				\draw[fill=green!20,color=pink!20] (-4,1.5) --++(8,0) --++ (0,1.75) --++(-8,0) -- cycle;
				\draw[fill=green!20,color=pink!20] (-4,1) --++(8,0) --++ (0,-1.75) --++(-8,0) -- cycle;
				\draw[fill=blue!20,color=blue!20] (-3,0) --++(0.5,-0.5) --++(5,0) --++(0.5,0.5) --++(-3,3) -- cycle;
				\draw[fill=green!20,color=green!20]  (-3,0) --++(0.5,-0.5) --++(5,0) --++(0.5,0.5) --++(-0.5,0.5)--++(-5,0)--cycle;
				\node at (0,0) {$L^-$};
				\draw[fill=green!20,color=green!20] (-1,2) --++(2,0) --++(-1,1)--cycle;
				\node at (0,2.5) {$L^+$};
				\draw[color=blue!20,thick] (-3,0) --++(0.5,-0.5) --++(5,0) --++(0.5,0.5) --++(-3,3) -- cycle;
				\node at (1.5,2) {$L$};
			\draw[fill=orange,color=orange] (-4,1)--++(8,0)--++(0,0.5) --++(-8,0) -- cycle;
			\node at (-3.5,1.25) {$S$};
			\draw[fill=orange,color=orange] (-2,1)--++(4,0)--++(-0.5,0.5) --++(-3,0) -- cycle;
			 {\draw[dotted] (-1,2) --++(1,-1); 
				\draw[dotted] (1,2) --++(-1,-1);
				\draw[fill=green!20] (-0.5,1.5) --++(0.5,-0.5) --++(0.5,0.5) --cycle;
				\node at (0,1.26) {$N$};}			
		\end{tikzpicture}
	\end{center}
	\caption{Set up for the localisation result of~\cite{FewsterJuarezAubry:202?}.\label{fig}}
\end{figure}

With this result in hand, we adapt Section~\ref{sec:noncompact} by changing the  external electromagnetic vector potential to $A = \xi \nabla\chi$, where $\xi$ is compactly supported and $\chi$ is as before. By arranging that $\xi=1$ on a neighbourhood of the closure of $L\cap S$, we can engineer similar results to those in~\eqref{eq:vareps_id} and \eqref{eq:nonsel_id} for observables localised in $L^+$. Here, $\Cf'$ is the coupled theory with noncompact coupling zone, while $\Cf$ will be the theory with the cut-off potential $\xi\nabla \chi$. In the $s=-\pi/2$ case, the new induced observable map obeys $\varepsilon_\sigma(B)=B$ for all probe observables localised in $L^+$. Therefore all such observables have a measurement scheme with compact coupling zone given by the support of $\xi$. In particular this applies to those localisable in the region $N$ within the coupling zone. As the whole set up can be designed for any region $N$ we conclude that all local observables of the Klein--Gordon field have an exact measurement scheme with compact coupling zone. This resolves Problem 1.

Similarly, using the same strategy with $s=\pi/2$, the nonselectively updated state agrees with the probe preparation state on observables in $L^+$, but agrees with the original state on the causal complement of $\supp\xi$, by general arguments given in~\cite{FewVer_QFLM:2018}. Going further, return to the situation of Alice and Bob in Problem 2. Because $O_A$ and $O_B$ are spacelike separated, we may design regions $L^\pm_{A/B}$ and $L_{A,B}$ with the properties above, so that $O_{A/B}=L^+_{A/B}$ and
$L_A$ and $L_B$ are spacelike separated. Alice and Bob each apply the state preparation protocol, each with their own probe and using their desired state as the probe preparation state. This can be regarded as a single nonselective update of $\omega$ using a `superprobe' comprising the probe theories of both Alice and Bob. On the three-way tensor product of system and the two probes, Alice's scattering map obeys $\Theta_A X\otimes\one\otimes\one = \one\otimes X\otimes \one$ for $X\in\Af(L_A^+)$, while Bob's obeys $\Theta_B Y\otimes\one\otimes\one = \one\otimes   \one\otimes Y$ for $Y\in\Af(L_B^+)$. Due to a property of \emph{causal factorisation}~\cite{FewVer_QFLM:2018}, the combined scattering map obeys $\Theta (XY)\otimes \one\otimes \one=\one\otimes X\otimes Y$ for  $X\in\Af(L_A^+)$, $Y\in\Af(L_B^+)$, and so the overall nonselectively updated state  obeys 
\begin{equation}	
	\omega_{\text{ns}}(XY) = (\omega\otimes\omega_A\otimes\omega_B)(\Theta (XY)\otimes \one\otimes\one) = (\omega\otimes\omega_A\otimes\omega_B)(\one\otimes X\otimes Y) =\omega_A(X)\omega_B(Y)
\end{equation}
for such $X,Y$, and is therefore a local product state of $\omega_A$ and $\omega_B$.
Finally, because $\omega_{\text{ns}}$ is obtained by nonselective update, we conclude that it is Hadamard, provided that $\omega$ and $\sigma_A$, $\sigma_B$ are. This concludes the positive resolution of Problem 2. 

%
%

%
%


\paragraph{Acknowledgments} This work is supported by EPSRC Grant EP/Y000099/1 to the University of York. I thank Benito Ju\'arez-Aubry for useful comments on the manuscript.

\providecommand{\newblock}{}


\begin{thebibliography}{10}
\bibitem{FewsterJuarezAubry:202?}
	Fewster C~J and Ju\'{a}rez-Aubry B How local need local measurements be? {I}n
	preparation
\end{thebibliography}
\end{document}